\documentclass[aps,pra,amsmath,superscriptaddress,floatfix,notitlepage,balancelastpage,twocolumn,showpacs]{revtex4}
\usepackage{graphicx,subfigure} 
\usepackage{amsmath} 
\usepackage{amsthm} 
\usepackage{amssymb}	

\usepackage{tikz}

\newcommand{\gv}[1]{\ensuremath{\mbox{\boldmath$ #1 $}}}
\newcommand{\abs}[1]{\left| #1 \right|} 
\newcommand{\avg}[1]{\left< #1 \right>} 
\newcommand{\f}[2]{\frac{#1}{#2}} 
\newcommand{\dd}[2]{\frac{d^2 #1}{d #2^2}} 

\let\baraccent=\= 
\renewcommand{\=}[1]{\stackrel{#1}{=}} 

\theoremstyle{definition}

\theoremstyle{remark}

\newcommand{\nb}{n^{}_{\rm B}}

\newcommand{\nsites}{N_{\rm sites}}

\newcommand{\as}{a_s}

\newcommand{\tc}{{T_{\rm c}}}

\newcommand{\curO}{{\cal O}}

\newcommand{\kb}{k_{\rm B}}

\newcommand{\bk}{{\bf k}}

\newcommand{\hh}{\hat{h}}

\newcommand{\br}{{\bf r}}
\newcommand{\bn}{{\bf n}}

\newcommand{\be}{\begin{equation}}
\newcommand{\ee}{\end{equation}}
\newcommand{\bea}{\begin{eqnarray}}
\newcommand{\eea}{\end{eqnarray}}
\newcommand{\bse}{\begin{subequations}}
\newcommand{\ese}{\end{subequations}}

\begin{document}

\title{A self-consistent Hartree-Fock approach for interacting bosons in optical lattices}
\author{Qin-Qin L\"u}
\affiliation{Department of Physics and Astronomy, Louisiana State University, Baton Rouge, LA 70803, USA}
\author{Kelly R. Patton}
\affiliation{School of Science and Technology, Georgia Gwinnett College, Lawrenceville, GA 30043, USA}
\author{Daniel E. Sheehy}
\email{sheehy@lsu.edu}
\affiliation{Department of Physics and Astronomy, Louisiana State University, Baton Rouge, Louisiana, 70803, USA}
\date{July 29, 2014}
\begin{abstract}  
A theoretical study of interacting bosons in a periodic optical lattice
is presented.
Instead of the 
commonly used tight-binding approach (applicable  near the Mott
insulating regime of the phase diagram), the present work
starts from the exact  single-particle states of bosons in a cubic optical
lattice, satisfying the Mathieu equation, an approach that can be particularly
useful at large boson fillings.   The effects of short-range interactions are 
incorporated 
using a self-consistent Hartree-Fock approximation, and predictions for 
experimental observables such as the superfluid transition temperature,
condensate fraction, and boson momentum distribution are presented.
\end{abstract}


\maketitle

\section{Introduction} Ultracold atoms in optical lattices have
recently emerged as a novel setting for physicists to study
interacting many-body systems~\cite{rmp:bloch,review:Lewenstein}. 
Usually made by a set of standing waves
that are formed by interfering counter-propagating laser
beams, optical lattices mimic the crystalline
lattice potential in condensed matter systems.

The single particle potential for bosons in an optical lattice can be taken 
to be a cosine function of position in each orthogonal direction.  At
sufficiently low temperatures, and for sufficiently large optical lattice
amplitude $V_0$, one can approximate such a system
by an effective boson Hubbard model (BHM), in which the minima of the 
single-particle potential correspond to sites of the Hubbard model~\cite{prl:jaksch1998}.
As first shown by Fisher et al.~\cite{prb:fisher1989}, the boson Hubbard model exhibits, at 
integer filling, a quantum phase transition between the superfluid phase
and an incompressible Mott insulating phase.  

Starting with the pioneering work of Greiner et al.~\cite{nat:greiner}, numerous experiments have explored
the properties of bosons in optical lattices that realize
the BHM~\cite{jlowtemp:kohl,prl:spielman2007,prl:MunKetterle,nat:chin,natphys:trotzky,njp:becker,sci:endres2011,prl:marknageri,sci:chin}.
  The transition from the superfluid to the Mott phase occurs with
increasing $U/J$, where $U$ and $J$ are the on-site repulsion and nearest-neighbor
tunneling matrix elements, respectively, in the BHM.  These phases are separated by a 
quantum critical point at which the BEC transition
temperature $\tc$ is suppressed to zero with increasing $U/J$. This suppression was observed
experimentally by Trotzky et al.~\cite{natphys:trotzky}, who could control $U/J$ by tuning the optical lattice
depth parameter $V_0$,
 quantitatively confirming the BHM picture for bosons
at unit filling.

The purpose of the present work is to explore bosons in optical lattices via a
different approach without making the simplification to the BHM Hamiltonian but,
rather, by studying the full Hamiltonian for bosons in a periodic optical
lattice potential with short-ranged interactions.  One motivation for our
study is the fact that even {\em non-interacting\/} bosons in a periodic optical
lattice will exhibit a strong suppression of the BEC
transition temperature $\tc$ with increasing $V_0$ (although $\tc$ will always be nonzero),
that is essentially due to the increasing effective mass (or flattened single-particle bands)
associated with a larger optical lattice amplitude.
  The question we pose, then, is to what extent the $\tc$ suppression observed by
Trotzky et al.~could be understood within this simple effective mass picture.

\begin{figure}[htb]
  \begin{center}
  \includegraphics[width=0.9\columnwidth]{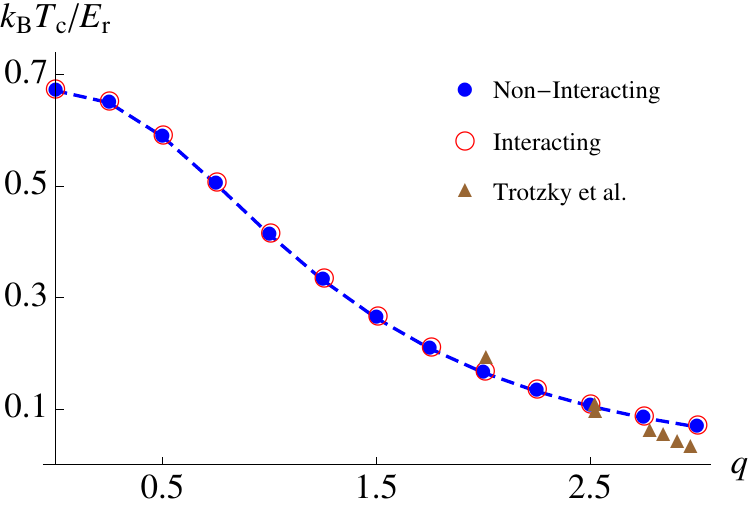}\\
  \end{center}
  \caption{ (Color online) We show in this plot the transition temperature $\tc$, normalized to
the recoil energy $E_{\rm r}$, as a function of the normalized optical lattice depth
$q=V_0/4E_{\rm r}$, for bosons at unit filling in a periodic optical lattice potential.
The blue points (and dashed curve) show the non-interacting case, the red circles show
our interacting Hartree-Fock calculation, and the triangles show the experimental data
from Ref.~\cite{natphys:trotzky} (indicated as ``Trotzky et al.").   The latter shows a clear suppression for larger $q$ as the Mott 
insulating quantum critical
point (at $q\simeq 3$ in this figure~\cite{Capogrosso}) is approached
}\label{fig:HFSCUnitFillingTcvsQ}
\end{figure}

More generally, we are interested in understanding how interaction effects impact
the observable properties of bosons in optical lattices far away from the regime
where the BHM applies at low temperatures and large optical lattice depth.  Our 
starting point is the problem of non-interacting bosons in a periodic potential.
As we discuss below, the corresponding single-particle problem that we need to 
solve to describe this system is the one-dimensional Schr\"odinger equation for bosons
in a cosine-shaped potential, also known as the Mathieu equation~\cite{Abramowitz}.  We note
that other recent theoretical works have explored the Mathieu equation in this context,
including Zwerger~\cite{Zwerger}, who used the known bandwidth of the Mathieu equation to
derive an approximation for the Hubbard tight-binding parameter, and McKay et al.~\cite{pra:demarco}, who studied
the thermodynamics of trapped cold bosons using the Mathieu equation.

An additional question of interest, motivating our work, is how short range repulsive interactions 
(characterized by scattering length $a_s>0$ or BHM repulsion $U>0$) impact
observable properties of bosons such as $\tc$.   For large optical lattice depth and low filling, where the BHM 
applies, increasing the strength of repulsive interactions suppresses $\tc$ as the Mott phase is approached.
In contrast, for a uniform BEC (equivalent to our system at optical lattice depth $V_0 =0$), 
increasing the repulsive interactions leads to an increase of $\tc$~\cite{Baym99,rmp:andersen}. 
To investigate this, we incorporate interactions for bosons in a periodic optical lattice within a self-consistent
Hartree-Fock  approximation.  While  Hartree-Fock is known to have a vanishing affect on $\tc$ for a uniform gas,
we find a small $\tc$ enhancement for increasing $a_s$ for bosons in a periodic optical lattice.  

Before proceeding to the details of our calculations, we first present our main results.  
In  Fig.~\ref{fig:HFSCUnitFillingTcvsQ} we show $\kb\tc$ (with $\kb$ the Boltzmann constant) for a non-interacting
BEC in a periodic optical lattice, normalized to the recoil energy $E_{\rm r} = \frac{\hbar^2 k^2}{2m}$, 
as a function of optical lattice depth $V_0$ in the combination $q \equiv V_0/4E_{\rm r}$, along with
the results of the Trotzky et al.~experiment and also our interacting Hartree-Fock approach (using the same
parameters as the Trotzky et al.~experiment).  Incorporating the Trotzky et al.~results into this figure 
required expressing the data of Ref.~\cite{natphys:trotzky} in terms of the parameters $ V_0/4E_{\rm r}$ via
an approximate tight-binding formula for the hopping matrix element $J$, as described below.  However, 
this plot shows that the Trotzky et al.~$\tc$ data quantitatively agrees with the non-interacting theory for
small optical lattice depth, and shows a clear suppression for larger optical lattice depth as the Mott insulating
quantum critical point is approached.
%

\begin{figure}[ht!]
     \begin{center}
        \subfigure{
            \includegraphics[width=85mm]{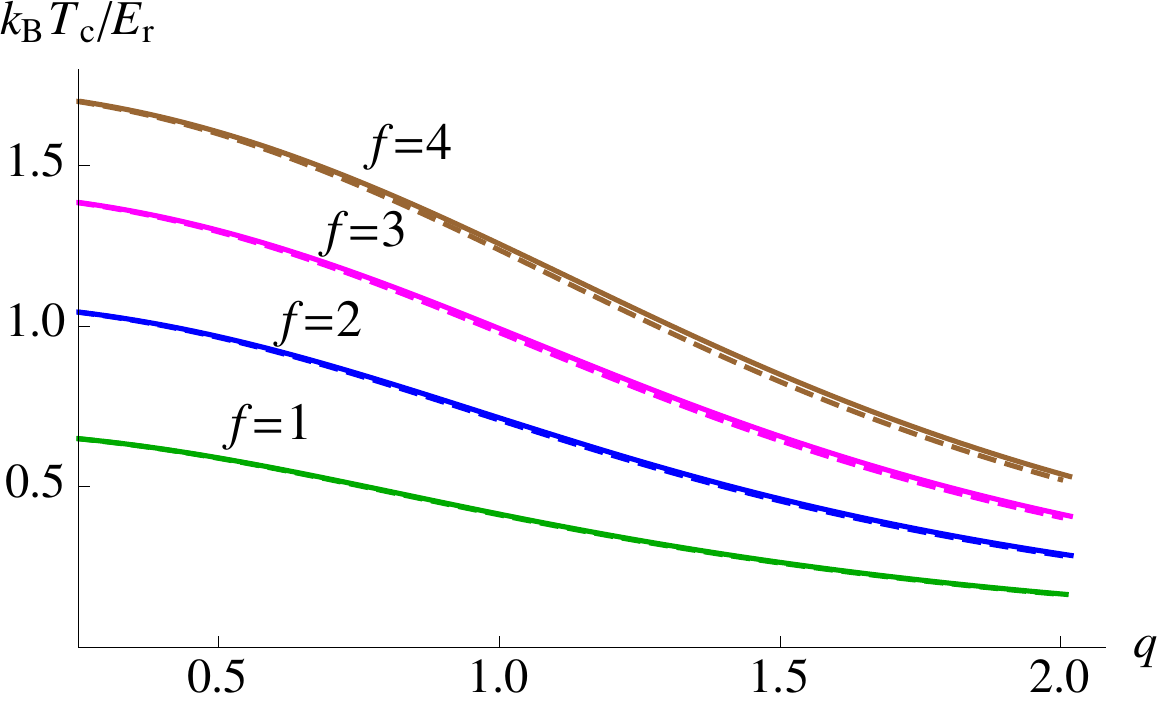}}
        \\ \vspace{-.25cm}
        \subfigure{
            \includegraphics[width=85mm]{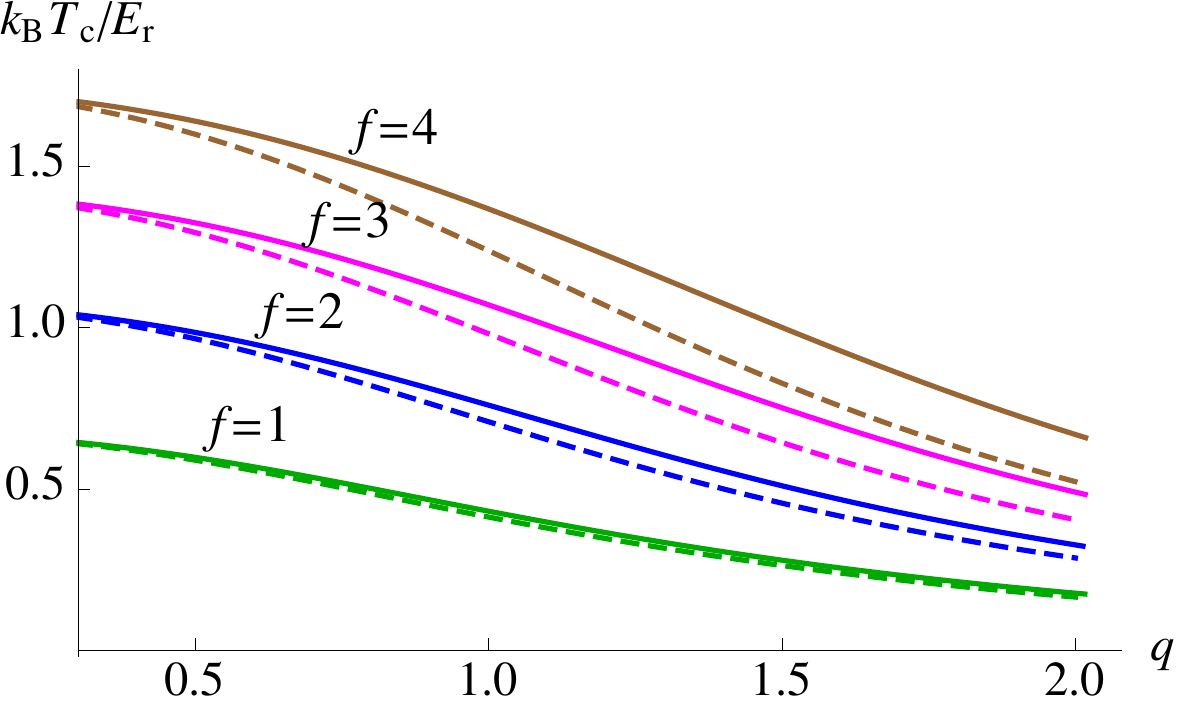}}
    \end{center}\vspace{-.5cm}
    \caption{(Color online) The top panel shows $\tc$ vs.~normalized optical lattice depth $q$ for
parameters consistent with the Trotzky et al.~experiment, but including larger filling values ($f = N/\nsites$
with $N$ the particle number and $\nsites$ the number of lattice sites).  For each case, the solid curve is
the interacting case and the dashed curve is the non-interacting case. 
Although these curves show a slight separation of the non-interacting and interacting curves with
increasing filling $f$, the difference is quite small even for the largest filling.  In the bottom
panel we plot $\tc$ vs.~$q$ for larger $\as$ ($a_s/a = 0.1$), which shows a significant enhancement of $\tc$ due to
interactions.
     }\vspace{-.25cm}
   \label{fig:TcvsQMultiFilling}
\end{figure}

Figure~\ref{fig:HFSCUnitFillingTcvsQ} also shows that our interacting Hartree-Fock approach is indistinguishable from non-interacting 
bosons in an optical lattice in this parameter regime (although our interacting $\tc$ is slightly higher
than the non-interacting case).  In Fig.~\ref{fig:TcvsQMultiFilling}, we show our results for various filling values at small scattering
length (top panel, $a_s$ consistent with parameters of Ref.~\cite{natphys:trotzky}) and for large scattering length 
(bottom panel $a_s/a = 0.1$, with $a$ the optical lattice spacing, but other parameters still consistent with 
 Ref.~\cite{natphys:trotzky}), with only the latter showing a significant enhancement of the transition temperature
arising from the repulsive interactions.  


Our work can be summarized as the following: we construct the wave
functions for bosons in an optical lattice using Mathieu Functions,
and obtain the single particle energies from the eigenvalues of
Mathieu equation. We are then able to calculate experimental
observables for bosons in an optical lattice and verify their
agreement with experiments. Within our Hartree-Fock self-consistent
scheme, we find that interaction raises the critical temperature,
makes more atoms condense,  and results in a more uniform boson
density. The finite size effect and boundary conditions are also
considered in our calculation.

We organize this paper as follows. In
Sec.~\ref{sec:mathieu} we will introduce the Mathieu equation which
naturally describes the single-particle states of non-interacting bosons in an
optical lattice, with the transition temperature and number equations 
depending on the Mathieu equation eigenvalues.  In Sec.~\ref{sec:HFAns}, we 
describe our method for 
incorporating interaction effects using a self-consistent Hartree-Fock approach
that leads to coupled equations that must be solved numerically.
In Sec.~\ref{sec:results}  we present our results from solving these equations,
and describe how repulsive interactions modify observables like the superfluid
transition temperature, condensate fraction, local boson density and 
boson momentum distribution.
 In this section, we initially choose
system parameters consistent with the experiments of Trotzky et al.~\cite{natphys:trotzky} before
subsequently considering the effect of larger filling and larger scattering length.  
 Section~\ref{sec:conclude} concludes the paper and
provides some additional discussion.

\section{System Hamiltonian and non-interacting limit}
\label{sec:mathieu}

Our model Hamiltonian $H = H_0 + H_1$ for bosons in an optical lattice consists of a single particle ($H_0$) and interaction ($H_1$) piece:
\bea
\label{Eq:hnought}
H_0 &=& \int d^3 r \, \Phi^\dagger(\br) \Big[ 
- \frac{\hbar^2\nabla^2}{2m} - \mu + V(\br)
\Big]\Phi(\br) ,
\\
H_1 &=&  \frac{g}{2} \int  d^3 r  \,\Phi^\dagger(\br) \Phi^\dagger(\br) \Phi(\br) \Phi(\br), 
\label{eq:hone}
\eea
where $\Phi(\br)$ is a bosonic field operator satisfying $[\Phi(\br),\Phi^\dagger(\br')] = \delta^{(3)}(\br-\br')$, $m$ is the boson mass, and $\hbar$ is Planck's constant.  Here, 
$g = \frac{4\pi \hbar^2 a_s}{m}$ with  $a_s$ the s-wave scattering length, and 
$V(\br) = V_0(\cos^2{kx}+\cos^2{ky}+\cos^2{kz}-\f{3}{2})$ is the imposed optical lattice potential characterized by the
optical lattice depth $V_0$ and the wavevector $k$ (with the lattice spacing $a= \pi/k$).
The subtracted constant 3/2 ensures the spatial integration of $V(\br)$ vanishes.

\begin{figure}[htb]
 \includegraphics[width=0.9\columnwidth]{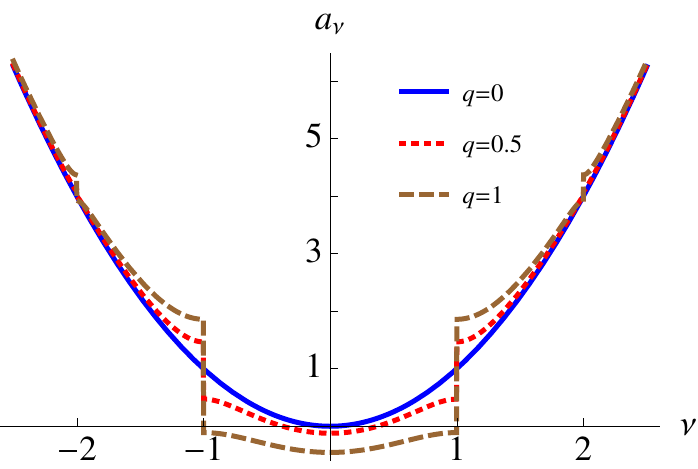}\\
  \caption{(Color Online) The Mathieu characteristic function for the even Mathieu
function gives the dispersion, i.e. $a_\nu=\varepsilon_\nu/E_{\rm r}$. Here $q=V_0/4E_{\rm r}=1$, $0.5$, $0$. The ground
state energy (the bottom of the curves) is lowered as the optical lattice potential $V_0$ increases.
The Characteristic Function is reduced to a parabola when $V_0=0$.}\label{fig:MathieuCharAmanyqMainText}
\end{figure}

In the absence of interactions, $g=0$, $H$ is solvable by considering the eigenfunctions of the single-particle Hamiltonian $\hh \equiv - \frac{\nabla^2}{2m} + V(\br)$, 
(henceforth we take $\hbar = 1$) that satisfy 
\be
\label{Eq:originaleigenproblem}
\hh\Phi_{\gv{\nu}} (\br)=E_{\gv{\nu}} \Phi_{\gv{\nu}} (\br),
\ee
where $\gv{\nu}=(\nu_x, \nu_y, \nu_z)$ is the eigenvalue index,
$E_{\gv{\nu}}$ is the total energy, and   $\Phi_{\gv{\nu}}(\br)$ is the 3D wave
function.  We can write  $\Phi_{\gv{\nu}}(\br)$ as a product of
wave functions in the $x$, $y$, and $z$ directions as  $\Phi_{\gv{\nu}}
(\br)=\phi_{\nu_x}(x)\phi_{\nu_y}(y)\phi_{\nu_z}(z)$, with each of the $\phi_{\nu_i}$ satisfying
a corresponding 1D Schr\"odinger equation with a 1D potential:
\be
\left[-\frac{1}{2m} \frac{d^2}{dx^2} + 
V_0\left(\cos^2{kx}-\frac{1}{2}\right)\right] \phi_\nu(x) =\varepsilon_\nu \phi_\nu(x),
\ee
with 1D eigenvalue $\varepsilon_\nu$.  This 
can furthermore be rearranged into the form of the Mathieu equation~\cite{Abramowitz}: 
\begin{equation}
\dd{\phi_\nu (u)}{u}+(a_\nu-2q\cos{2u})\phi_\nu(u)=0,
\label{equ:mathieu}
\end{equation}
where $u = kx$ is a dimensionless coordinate.  Here, $a_\nu = \frac{ \varepsilon_\nu}{E_{\rm r}}$ and $q = \frac{V_0}{4E_{\rm r}}$
are dimensionless forms of the 1D eigenvalue and optical lattice depth, normalized to the recoil energy $E_{\rm r} = \frac{k^2}{2m}$.
The Mathieu Equation (Eq. $\ref{equ:mathieu}$) has even and odd periodic solutions, $ce(a_\nu,q,u)$ and $se(b_\nu,q,u)$, respectively,
with $a_\nu$ and $b_\nu$ called the Mathieu characteristic functions (playing the role of the eigenvalue here) for the even and odd
solutions.  The real number $\nu$ determines the periodicity of the solutions, and generally $a_\nu = b_\nu$ except when $\nu$ is 
an integer.  In Fig.~\ref{fig:MathieuCharAmanyqMainText}, we plot $a_\nu(q)$ as a function of $\nu$ for three values of the
normalized optical lattice depth, $q=0,~0.5,~1$, showing a typical band structure for particles in a periodic potential, with $\nu =1$
being the Brillouin zone boundary.

 The Mathieu equations solutions $ce(a_\nu,q,u)$ and $se(b_\nu,q,u)$,  analogous to cosine and sine, respectively,
 can also be combined into analogues of complex exponential functions as:
\be
me_\nu(u,q) = ce(a_\nu,q,u)+i\, se(b_\nu,q,u),
\ee
which satisfy a Bloch theorem:
\be
\label{eq:mebloch}
me_\nu(u+n\pi,q) ={\rm e}^{in\pi \nu}  me_\nu(u,q).
\ee
Here,  $n$ is any integer, so that $\nu$ can be regarded as a Bloch quasi-momentum, with $p= \pi \nu/a$.  

 To study the BEC, we consider a box of volume $V = L^3$ that encloses
$N_{\rm s}$ lattice sites along each direction, with $\nsites = 
N_{\rm s}^3$ being the total number of lattice sites in the cubic lattice.  Imposing periodic 
boundary conditions implies, for our 1D solutions, 
$\phi_\nu(k[x+L])=\phi_\nu(kx)$.   Using Eq.~(\ref{eq:mebloch}) with $n\pi = kL =\pi N_{\rm s}$, we have 
\be
me_\nu(u+kL,q) ={\rm e}^{i\pi \nu N_s}  me_\nu(u,q),
\ee
which implies the $\nu$ satisfy $\nu_m = 2m/N_{\rm s}$ with $m$ any integer, to have the phase on the right side be unity.  Therefore, our
quantized wave functions for bosons in an optical lattice with periodic boundary conditions can be written as 
\be
\phi_{\nu_n}(x) = \begin{cases} \frac{1}{\sqrt{L}} me_{\nu_n}\Big(kx,q\Big),& \text{if $n \neq 0$ },\cr
 \sqrt{\frac{2}{L}} ce\Big(a_0 ,q,kx\Big), & \text{if $n=0$ },
\end{cases} 
\label{Eq:onedwf}
\ee
where the special case of $n=0$ occurs because the odd Mathieu function is not defined for $\nu=0$.  With this definition, the $\phi_{\nu_n}(x)$
satisfy the normalization
\be
\int_0^L dx \phi_{\nu_n}^*(x) \phi_{\nu_m}(x) = \delta_{mn}.
\ee
The particle number equation used to determine the BEC transition temperature $\tc$ and condensate fraction below $\tc$ is
\be
\label{Eq:numberequation}
 N = N_0 + \sum_{\bn \neq 0} \nb\big( E_{\bn} - \mu  \big) ,
\ee
with $N$ the total particle number, $N_0$ the number in the lowest state $\Phi_{\bf 0}(\br) = \phi_0(x)\phi_0(y)\phi_0(z)$, 
and $E_\bn = E_{\rm r}(a_{\nu_x} + a_{\nu_y}+ a_{\nu_z})$.    The sum in Eq.~(\ref{Eq:numberequation}) is understood to be over integers
$n_x$, $n_y$, and $n_z$ from $-\infty$ to $\infty$.  Approximating the sum by an integral by introducing the continuous variable
$\nu_x = 2n_x/N_{\rm s}$ (and similarly for $\nu_y$ and $\nu_z$), we have  
\bea
\nonumber
N&=& N_0 + \frac{N_{\rm s}^3}{8} \int_{-\infty}^\infty d^{3}\nu\,
\nb\big( E_{\gv{\nu}} - \mu  \big) ,
\\
&=& N_0 + \nsites \int_{0}^\infty d^{3}\nu\,  \nb\big( E_{\gv{\nu}} - \mu  \big) ,
\label{eq:neq}
\eea
where in the second line we used the symmetry of the integrand under $\gv{\nu} \to - \gv{\nu}$ to simplify the integrals and 
introduced $\nsites  = N_{\rm s}^3$, the total number of lattice sites in our system.

 We can then solve for the superfluid transition temperature
$\tc$ from Eq.~(\ref{Eq:numberequation}), which occurs when the
chemical potential reaches the lowest state, i.e. $\mu/E_{\rm r}=
3a_{0}(q)$, with $a_{0}(q)$ referring to the characteristic function's
minimum at $\nu=0$.   As usual for a BEC, the condensate number
below $\tc$ is determined by Eq.~(\ref{eq:neq}) with $\mu$ pinned
to the bottom of the band ($\mu = 0$ for a free gas, but $\mu = E_0$ for the present case).  Having established the notation of the
Mathieu equation and reviewed the non-interacting BEC problem for 
this case, we now turn to the interacting case and present our self-consistent Hartree-Fock approach.

\section{Hartree-Fock Self-Consistent Scheme: Ansatz}
\label{sec:HFAns}

In the preceding section, we studied
non-interacting bosons in an optical lattice with the Mathieu equation. In
this section, we try to capture interaction effects by using
Hartree-Fock approximation. For bosons in a uniform potential,
interaction effects vanish identically within the Hartree-Fock
approximation~\cite{Baym99}.  This follows because, for a 
uniform gas, the Hartree-Fock contribution to interactions enter as
a shift in the chemical potential $\mu\to \mu - 2gn$ with $n$ the 
local density, and can therefore be absorbed in a redefinition of the
chemical potential. 

 In the presence of an optical lattice potential,
this translational invariance is broken and physical properties 
such as the superfluid transition temperature can be modified by
interaction effects, even within the Hartree-Fock approximation.  This is seen most strikingly in the suppression
of the transition temperature to $0\, {\rm K}$ for bosons at integer filling,
resulting in a quantum phase transition to the  Mott insulating
state.  Here our main interest is studying such interaction effects
away from the Mott regime at low temperature and integer filling, using 
a self-consistent Hartree-Fock approach that utilizes the Mathieu 
function representation for bosons in an effective periodic potential.

Our self-consistent
Hartree-Fock approximation is motivated by first noting that  bosons
in a periodic cosine-shaped potential will have a local density that
is also periodic. Approximately, this density is given by a constant piece plus
a spatially-modulated cosine-shaped piece. Within the simplest Hartree-Fock 
approximation, one makes the replacement, for $H_1$,
\bea
\int \!\! d^3 r \Phi^\dagger(\br) \Phi^\dagger(\br) \Phi(\br) \Phi(\br) \!\!\!&\to& \!\!\!
2\int \!\! d^3 r \Phi^\dagger(\br) \Phi(\br)  \langle \Phi^\dagger(\br) \Phi(\br) \rangle ,
\nonumber
\\ 
&=&2\int  d^3 r \Phi^\dagger(\br) \Phi(\br) n(\br),
\label{Eq:scheme}
\eea
with the $2$ coming from the two ways such a contraction can occur, 
so that a spatially-periodic boson density $n(\br)$ acts like an additional single-particle
potential $\propto gn(\br)$ on the bosons.

Although Eq.~(\ref{Eq:scheme}) contains the essential physics of our scheme, we now derive it
via a more formal method.  To do this we consider the single-particle
Green's function for bosons described by the Hamiltonian $H$:
\be
G(\br_1,\tau_1;\br_2,\tau_2) = - \langle T_\tau \Phi(\br_1,\tau_1) \Phi^\dagger(\br_2,\tau_2) \rangle , 
\label{eq:greendef}
\ee
where $\tau$ refers to imaginary time, $T_\tau$ is the imaginary time ordering operator, and the time dependence of
$\Phi(\br,\tau)$ is determined by the Heisenberg equation of motion 
\be
\frac{\partial \Phi(\br,\tau)}{\partial \tau} = \big[H,\Phi(\br,\tau)\big].
\ee
Because our system is translationally invariant in the time direction, $G(\br_1,\tau_1;\br_2,\tau_2)$ can be taken to be a function
only of $\tau_1-\tau_2$ and furthermore can be expressed in terms of a sum over bosonic Matsubara frequencies:
\be
\label{eq:greenfourier}
G(\br_1,\tau_1;\br_2,\tau_2) = k^{}_{\rm B}T\sum_\omega {\rm e}^{-i\omega(\tau_1-\tau_2)}G(\br_1,\br_2;\omega).
\ee
The Dyson equation for $G(\br,\br';\omega)$ is:  
\bea
\label{eq:selfform}
&&G(\br,\br';\omega)  = G_0(\br,\br';\omega)
\\
&&\nonumber
 + \int d^3 r_{1} \,d^3 r_{2}\, G_0(\br,{\bf r}_{1};\omega) \Sigma({\bf r}_{1},{\bf r}_{2};\omega)G({\bf r}_{2}, \br';\omega), 
\eea
with
$\hh(\br) \equiv - \frac{\nabla^2}{2m} - \mu + V(\br)$.  Here,
$G_0(\br,\br';\omega)$ is the bare Green's function (for $H_1 = 0$) satisfying
\be
\label{eq:baregreen}
\big[i\omega - \hh(\br)\big]G_0(\br,\br';\omega)  = \delta(\br-\br'),
\ee
and
 $\Sigma({\bf r},{\bf r}';\omega)$ is the self-energy which, within
the Hartree-Fock approximation, has the form (as reviewed in Appendix~\ref{appnd:HF}):
\be
\label{Eq:sigmahf1}
\Sigma({\bf r},{\bf r}';\omega) = 2gn({\bf r}) \delta({\bf r}- {\bf r}') .
\ee
Plugging this into Eq.~(\ref{eq:selfform}), and acting on both sides with the operator $i\omega - \hh(\br)$,
we arrive at:
\be
\big[i\omega - \hh(\br)-2gn(\br) \big] G(\br,\br';\omega) =\delta(\br-\br'),
\ee
equivalent to:
\be
 G^{-1}(\br,\br';\omega)  = i\omega - \Big(-\frac{ \nabla^2}{2m}-\mu +V_0(\br) +2gn(\br)\Big) .
\label{Eq:greenhf}
\ee
so that, indeed, the Green's function within the Hartree-Fock approximation only depends on the
effective potential $V_0(\br) +2gn(\br)$.

Since the boson density $n(\br)$ is highest at minima of $V_0(\br)$, and because 
$g>0$,  the spatially-varying
part of $n(\br)$ will tend to cancel out the imposed periodic potential, so that the bosons
effectively \lq\lq see\rq\rq\ a lower lattice depth.
  As we shall see, this will tend 
to increase the transition temperature, and also make the BEC phase occurring below $\tc$ 
more spatially uniform than predicted by a non-interacting theory.  

To show this in detail, we proceed by making one additional approximation, by assuming
that the boson density as a function of position can be taken to be a constant piece
plus a piece that varies, spatially, in the same manner as the imposed optical lattice potential
[i.e., according to the function $v(\br)$]:
\be
\label{eq:ennapprox}
n(\br)\approx \frac{f}{a^3} [1-c v(\br)],
\ee
with $f=N/\nsites$ the filling,
 $v(\br) = \cos^2 k x +\cos^2 k y+\cos^2 k z - \frac{3}{2}$
the function appearing in the definition of the optical lattice potential,
and  $c$ an unknown parameter to be determined self-consistently.
   The approximation Eq.~(\ref{eq:ennapprox})
ensures $\int d^3 r \,n(\br)=N$, 
since the integral of the spatially dependent term over the unit cell vanishes. Because $\abs{v(\br)}<3/2$, for the density $n(\br)$ to 
be positive we need $-2/3<c<2/3$.   Additionally, 
since we expect the boson density to reach maxima at the minima of the lattice, we must have $c>0$.

\begin{figure}[htb]
 \includegraphics[width=\columnwidth]{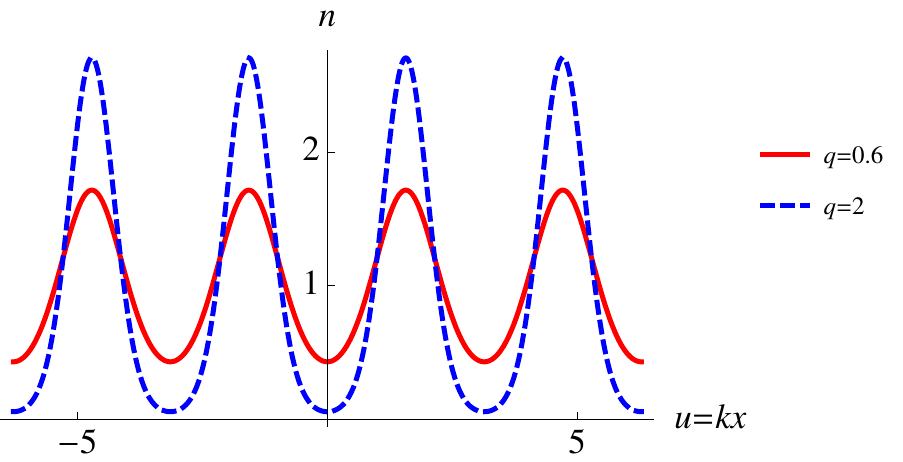}\\
  \caption{(Color online) Two calculated non-interacting boson  densities are compared in this plot. Here the dashed blue and solid red curves are
the modulus squared of the condensate wavefunction, $n=|\abs{me(0,q,u)}|^2$ for $q=2$ and $q=0.6$, respectively. When $q$ is small, the density
resembles a cosine shape plus a constant. However, the $q=2$ curves show deviations from this.  }
\label{fig:MathieuComparison}
\end{figure}

Translational symmetry dictates that $n(\br)$ have the same
periodicity as the lattice, so that $n(\br)$ has the same shape in
each unit cell.  However,  Eq.~(\ref{eq:ennapprox}) makes the
additional assumption that the spatial variation of $n(\br)$ is of the
same form as the imposed optical lattice, up to a scaling parameter,
which is $c$.  This assumption is valid at small optical lattice depth
since the modulus squared of the Mathieu functions is indeed
approximately given by a constant plus a cosine at small $q$, as
follows from the  expansion of Mathieu functions for small $q$
(Ref.~\cite{web:nistmath}):
\bea
&&me(\nu,q,u)=e^{i\nu u}-\frac{q}{4}\Big[\f{1}{\nu+1}e^{i(\nu+2) u}-\f{1}{\nu-1} e^{i(\nu-2) u}\Big]
\nonumber \\
&&
\qquad \qquad 
+\curO(q^2),
\label{equ:MathieuSmallqexpansion}
\eea
 where $me(\nu,q,u)=ce(a_\nu,q,u)+ise(b_\nu,q,u)$. This formula implies that if the normalized
lattice depth $q$ is sufficiently small, the only terms that will contribute
are the first line of Eq.~(\ref{equ:MathieuSmallqexpansion}).  To illustrate this, in Fig.~\ref{fig:MathieuComparison}  we plot
the modulus squared of the Mathieu functions for $q=0.6$ and $q=2$. While the $q= 0.6$ curve is clearly given by a constant plus
plus cosine piece, the $q =2$ curve exhibits deviations from this.  In the following, we aim to use Eq.~(\ref{eq:ennapprox})
beyond the small-$q$ regime. This amounts to assuming that the effect on interactions of the higher-order terms 
in Eq.~(\ref{equ:MathieuSmallqexpansion}) is small.

Within the preceding assumptions, the interacting Green's function Eq.~(\ref{Eq:greenhf})
 can be written as a sum over eigenfunctions of an effective Mathieu equation eigenproblem with a modified single-particle potential:
\begin{equation}
G(\br,\br';\omega) = \sum_{\gv{\nu}}\frac{\Phi_{\gv{\nu}}(\br)\Phi_{\gv{\nu}}(\br')}{i\omega -\epsilon^{}_{\gv{\nu}}}, 
\end{equation}
where the single-particle states $\Phi^{}_{\gv{\nu}}(\br)$ now satisfy 
\begin{equation}
\left[-\f{\nabla^2}{2m}+\left(V_0-\f{2gfc}{a^3}\right)v(\br)\right]\Phi^{}_{\gv{\nu}}(\br)=E^{}_{\gv{\nu}}\Phi^{}_{\gv{\nu}}(\br),
\label{equ:HFConsistMathieu}
\end{equation}
with $\epsilon^{}_{\gv{\nu}}=E^{}_{\gv{\nu}}+\f{2gf}{a^3}$. Thus, the eigenvalue problem Eq.~(\ref{equ:HFConsistMathieu}) is identical to the 
original eigenvalue problem Eq.~(\ref{Eq:originaleigenproblem}) but with a modified effective optical lattice potential 
$V_0-\f{2gfc}{a^3}$.  This implies that the solutions to Eq.~(\ref{equ:HFConsistMathieu}) are once again built from a product
of three Mathieu functions but with the replacement $q\to \bar{q}$ with 
\be
\bar{q}=\f{1}{4 E_{\rm r}}\left[V_0-\f{2gfc}{a^3}\right]=q-\f{4}{\pi}\f{a_s}{a}fc,
\label{equ:EffectiveLatDepthHFConsis}
\ee
where in the second equality we used the relation $g=4\pi  a_s/m$ between the coupling parameter and the boson $s$-wave scattering length
$a_s$.  Thus, as noted above, that interaction effects can be seen as canceling part of the optical lattice, since $\bar{q}<q$.
The unknown parameter $c$ will be determined self-consistently by considering the thermodynamic equations of motion 
for our system.

\subsubsection{At the transition temperature}
We start by describing our self consistent scheme at the transition temperature $\tc$.  Since our system is effectively 
non-interacting within the Hartree-Fock approximation, the boson density at temperature $T\geq \tc$ is:
\be
n(\mathbf{r})=\sum_{\gv{\nu}} \nb(E_{\gv{\nu}}-\mu+\f{2gf}{a^3}) \abs{\Phi_{\gv{\nu}}(\mathbf{r})}^2,
\label{equ:densityHFConsist}
\ee
where $\nb$ is the Bose function.  For our ansatz to be sensible, Eq.~(\ref{equ:densityHFConsist}) should be equal to
Eq.~(\ref{eq:ennapprox}).  At $\tc$, such an agreement implies 
\be
\sum_{\gv{\nu}} \nb(E_{\gv{\nu}}-E_0)\abs{\Phi_{\gv{\nu}}(\mathbf{r})}^2=\f{f}{a^3}[1-c v(\br)],
\label{eq:HFSCdensityTc} 
\ee 
where on the left side we used that, at the transition temperature, the chemical
potential $\mu$ reaches the bottom of the effective dispersion 
that is the argument of the Bose function in Eq.~(\ref{equ:densityHFConsist}): 
$\mu = E_{0}+\f{2gf}{a^3}$.

In the following, we only impose Eq.~(\ref{eq:HFSCdensityTc}) in an average sense. 
To do this, we   consider the spatial integration
of Eq.~(\ref{eq:HFSCdensityTc}) over a unit cell. The right hand
side is $f$, since $v(\br)$ vanishes from spatial
averaging. Converting the summation $\sum_{\gv{\nu}} $ into an integral on 
the left hand side,
 we arrive at
\bea
\label{equ:NumberIntegralHFConsistent}
&& f= \f{{N_{\rm s}}^3}{V}\int d^3\nu~\int_{\rm cell} d^3r \,\nb(E_{\gv{\nu}}-E_0)
\\
&&
\times
\abs{me({2\nu_x},\bar{q},kx)}^2 \abs{me({2\nu_y},\bar{q},ky)}^2 \abs{me({2\nu_z},\bar{q},kz)}^2.
\nonumber 
\eea
Using the normalization of the Mathieu functions
\be
\int_0^\pi du~ \abs{me(2\nu_x, \bar{q}, u)}^2=\pi
\ee
and ${N_{\rm s}}^3/V=a^{-3}$, Eq.~(\ref{equ:NumberIntegralHFConsistent}) is reduced to
\be
\int d^3\nu~ \nb(E_{\gv{\nu}}-E_0)=f,
\label{equ:NumberEqHFConsis} 
\ee 
which ensures
Eq.~(\ref{eq:HFSCdensityTc}) holds, on average, in each unit cell. 

Next, we demand that Eq.~(\ref{eq:HFSCdensityTc}) holds for the
leading non-uniformity of the local density in each unit cell. To do
this, we multiply both sides of Eq.~(\ref{eq:HFSCdensityTc}) by
$v(\br)$ and integrate over unit cell, obtaining a second
self-consistent condition:  
\be \int_{\rm cell} d^3r ~n(\br) v(\br)=
-\f{3}{8} cf.  
\ee 
The left side of this equation can be simplified by introducing the function 
\be
I(\nu,\bar{q})=\int_0^1 d\ell  \abs{me({2\nu},\bar{q},\pi \ell)}^2\left(\cos^2{\pi \ell}-\f{1}{2}\right),
\ee
leading to:
\be
-\f{3}{8} cf=\int d^3\nu~ \nb(E_{\gv{\nu}}-E_0) \left[I(\nu_x, \bar{q})+I(\nu_y, \bar{q})+I(\nu_z, \bar{q})\right].
\label{equ:nvIntegralHFConsis}
\ee
From Eq.~(\ref{equ:nvIntegralHFConsis}) and
Eq.~(\ref{equ:EffectiveLatDepthHFConsis}), the effective lattice depth
$\bar{q}$ can be solved. Because the filling is known from the number
equation Eq.~(\ref{equ:NumberEqHFConsis}), the parameter $c$ can also
be obtained. From these results, the interaction effect on the
system's density profile is described.  Next, we explain how the same
scheme works for the non-superfluid phase above $\tc$ and in the
superfluid phase below $\tc$.

\subsubsection{Above the Transition Temperature} The effective lattice
depth $\bar{q}$ and the parameter $c$ are temperature dependent,
following from the fact that interaction effects depend on the density
distribution which is temperature dependent. When the system is above
$\tc$, all particles in the system are thermal, and the chemical
potential is no longer pinned at the bottom of the band. Therefore the
self-consistent conditions are Eq.~(\ref{equ:NumberEqHFConsis}) and
Eq.~(\ref{equ:nvIntegralHFConsis}), with $E_0$ replaced by
$\mu-\f{2gf}{a^3}$.

To obtain $\bar{q}$ and $c$, we can first obtain the filling $f$ from
 Eq.~(\ref{equ:NumberEqHFConsis}) for the critical
temperature, then solve for the chemical potential $\mu$ and $c$ from
the self-consistent conditions for $T>\tc$, which paves the way for us
to describe the spatial and thermodynamical properties of this
interacting system.

\subsubsection{Below the Transition Temperature}
When the system is below $\tc$, our self-consistent formulas are very 
similar, except that some of the bosons are in the condensate, 
and the chemical potential is once again pinned to the bottom of the effective
dispersion.  We have for the density: 
\be
n(\mathbf{r})=N_0\abs{\Phi_0(\br)}^2+\sum_{\gv{\nu}\neq 0} \nb(E_{\gv{\nu}}-\mu+\f{16}{\pi}\f{a_s}{a} f E_{\rm r}) \abs{\Phi_{\gv{\nu}}(\mathbf{r})}^2,
\label{equ:numberEqbelowTc}
\ee
where $N_0$ is the number of condensed particles and $\Phi_0(\br)$ is the ground-state wave function (a product of Mathieu functions 
for the $x$, $y$, and $z$ directions).

 Integrating each term in Eq.~(\ref{equ:numberEqbelowTc}) over a unit cell, we have
\be
f=\f{N_0}{N_{\rm sites}}+\int d^3\nu~ \nb(E_{\gv{\nu}}-E_0),
\label{equ:fillingEqbelowTc}
\ee
the generalization of Eq.~(\ref{equ:NumberEqHFConsis}) below $\tc$.  
 Similarly, by multiplying each term in Eq.~(\ref{equ:numberEqbelowTc}) by $v(\br)$ and integrating over 
the unit cell, we obtain:
\be
\begin{split}
-\f{3}{8} cf&=N_0\int_{\rm cell}d^3r ~\abs{\Phi_0(\br)}^2 v(\br)\\
&+\int d^3\nu~ \nb(E_{\gv{\nu}}-E_0) [I(\nu_x, \bar{q})+I(\nu_y, \bar{q})+I(\nu_z, \bar{q})].
\end{split} 
\ee
Using the definition of $\Phi_0(\br)$, we can rewrite
the first term as $3\f{N_0}{N_{\rm s}}  I_0(\bar{q})$, where
$I_0(\bar{q})\equiv2\int_0^1 d\ell \abs{ce(a_0, \bar{q},\ell)}^2 (\cos^2
\pi \ell-\f{1}{2})$. Then, combining with
Eqs.~$\eqref{equ:EffectiveLatDepthHFConsis}$ and
$\eqref{equ:fillingEqbelowTc}$, we are able to solve for $\bar{q}$ and
$c$ for the interacting system below $\tc$.

\section{Results}
\label{sec:results} In the previous section we described our Hartree-Fock self-consistent approach for interacting bosons in optical
lattices, in which the effect of inter-atomic interactions amounts to
an  effective periodic potential that partially offsets the imposed optical
lattice.  In this section, we present our numerical solution of the resulting
equations in several parameter regimes.  We will be interested in the 
shift of the transition temperature due to repulsive interactions, an issue that
has been pursued theoretically for decades in the case of a homogeneous
boson gas, with contradicting results including
both positive and negative $\tc$ shifts~\cite{rmp:andersen}.  

We find 
a small increase of $\tc$ with increasing repulsion within the self-consistent Hartree-Fock approximation that
we interpret, physically, as being due to
 a spatial homogenization of the local
boson density (relative to the non-interacting case) that makes it more likely for bosons to 
exchange with their neighbors, enhancing $\tc$.
 We also compute the condensate fraction below the transition temperature as well as additional observables,
such as the local boson density in a unit cell and the boson momentum distribution (measurable
via time of flight experiements), which reflect the predicted homogenization of the local boson density
in an optical lattice.

\subsection{Low filling and small scattering length}
We start with the case of
$^{87}$Rb atoms in an optical lattice with parameters 
consistent with the experiments of Trotzky et al.~\cite{natphys:trotzky}, before considering
larger filling and larger scattering lengths in subsequent sections.  Trotzky et al.~\cite{natphys:trotzky}
observed a suppression of the transition temperature $\tc$ for bosons at unit filling with increasing
$U/J$ (with $U$ the on-site repulsion and $J$ the nearest neighbor hopping matrix element)
that is quantitatively consistent with the presence of a quantum phase transition to the Mott insulating
state at $U/J \simeq 29.3$~\cite{Capogrosso}.
Figure 5 of Ref.~\cite{natphys:trotzky} shows experimental results, plotted as $\kb\tc/J$ vs. $U/J$.  To
compare to our theoretical calculations, we converted these data to the dimensionless parameters of our theory,  
$\kb\tc/E_{\rm r}$ and $V_0/E_{\rm r}$, using the approximate formulas~\cite{Zwerger,rmp:bloch} 
\bea
J&=& \frac{4}{\sqrt{\pi}} E_{\rm r} \left(\frac{V_0}{E_{\rm r}}\right)^{3/4}\exp\left[-2\left(\frac{V_0}{E_{\rm r}}\right)^{1/2}\right],
\\
U&=& \sqrt{\frac{8}{\pi}} k a_s E_{\rm r} \left(\frac{V_0}{E_{\rm r}}\right)^{3/4},
\eea
for the Bose Hubbard model parameters $J$ and $U$.  These can be combined to give:
\be
\label{Eq:combined}
\frac{V_0}{E_{\rm r}} = \frac{1}{4}\left[\ln \left( \frac{k a_s}{\sqrt{2}} \frac{J}{U}\right)\right]^2,
\ee
which we use with parameters consistent with Ref.~\cite{natphys:trotzky}, with $a_s = 5.31\,{\rm nm}$ for the
scattering length.  Although the optical lattice of Ref.~\cite{natphys:trotzky} is not quite cubic, 
with wavelength $\lambda_x = 765\,{\rm nm}$ and $\lambda_y = \lambda_z = 844\,{\rm nm}$, for simplicity we neglected 
this difference and used $k = 2\pi/844\, {\rm nm}$.  

As we have already discussed,  Fig.~\ref{fig:HFSCUnitFillingTcvsQ} shows 
$\tc$ within our self-consistent theory in comparison with the
Trotzky et al. data (using the abovementioned conversion) and in comparison with non-interacting bosons 
in a periodic optical lattice.
Thus, we see that the interacting Hartree Fock and non-interacting theories are indistinguishable.
This is expected, since Hartree-Fock type interaction effects are small at such low fillings.  Both theory
curves agree well with the Trotzky data at lower $q$ (suggesting that interaction effects are negligible
here), only disagreeing at large $q$, where the Trotsky et al.~data shows a clear
suppression towards the expected quantum critical point at $U/J = 29.3$~\cite{Capogrosso}.  Using 
Eq.~(\ref{Eq:combined}), this should occur at $q \simeq 3$.

\begin{figure}[htb]
  \begin{center}
  \includegraphics[width=0.87\columnwidth]{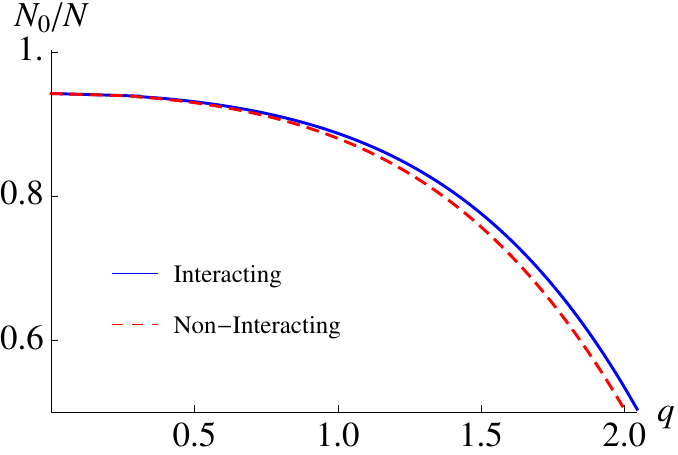}\\
  \end{center}
  \caption{(Color online) In this plot, we show the condensate fraction $N_0/N$
for both interacting and non-interacting gases, as a function of the
lattice depth $q=V_0/4E_{\rm r}$ at temperature $T=0.1E_{\rm r}$. The condensate fraction decreases with
increasing lattice depth $q$, approaching the phase transition. We also
observe that the condensation fraction for the interacting gas is
larger than that of the non-interacting gas. Here the system is at unit
filling, with system parameters consistent with those of
Ref.~\cite{natphys:trotzky}. }
\label{fig:HFSCFilling1n0vsQ}
\end{figure}

Within our theory, interaction effects become stronger below $\tc$ in the superfluid phase, as the condensate becomes occupied.
However, for system parameters consistent with Ref.~\cite{natphys:trotzky}, we still find interaction effects to
be small.  This is illustrated in Fig.~\ref{fig:HFSCFilling1n0vsQ},
which shows our theoretical prediction for the condensate fraction, $N_0/N$ vs. 
normalized optical lattice depth, $q = V_0/4E_{\rm r}$ for the case of $T = 0.1E_{\rm r}$, along with the case of 
vanishing interactions for comparison.  We see that interaction 
effects are small for any $q$, but are smallest for $q\to 0$ (the case of no optical lattice) with the interacting
condensate fraction being slightly larger than the non-interacting case at larger $q$.

Within our scheme, we do not expect to be able to capture the suppression of $\tc$ towards the Mott
phase and, as we have seen, we also find negligible effects of interactions away from the deep Mott
insulator regime for system parameters consistent with the Trotzky et al.~results.  Next, we turn
to the case of larger filling and larger interaction strength, where interaction effects may be
more significant.

\subsection{Interaction effects at large filling}

\begin{figure}[htb]
  \begin{center}
  \includegraphics[width=0.85\columnwidth]{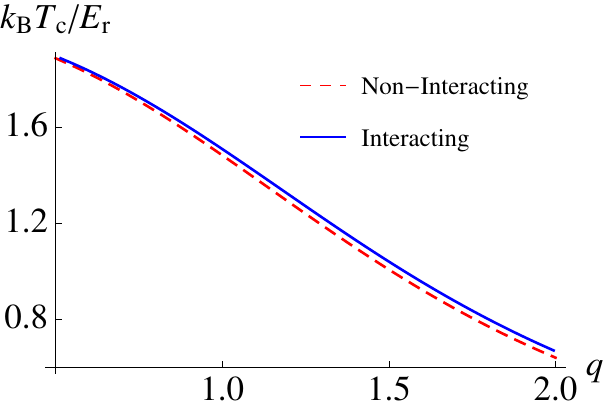}\\
  \end{center}
  \caption{ (Color online) In this plot, we show the trend of the
transition temperature $\tc$ with varying lattice depth
$q=V_0/4E_{\rm r}$ for the case of filling $f=5$ and other system parameters consistent with Ref.~\cite{natphys:trotzky}.}
\label{fig:HFSCFilling5TcvsQ}
\end{figure}

\begin{figure}[htb]
  \begin{center}
  \includegraphics[width=0.9\columnwidth]{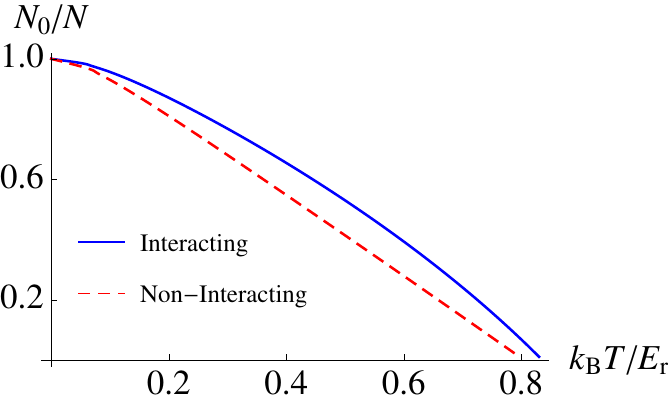}\\
  \end{center}
  \caption{(Color online) Plot of the condensate fraction for both non-interacting and
interacting gases, as a function of temperature, over the
temperature range from $T=0$ to $\tc$ (where the fraction becomes
zero), with filling $f=5$ and normalized optical lattice depth $q=1.75$,
 and other system parameters consistent with 
Ref.~\cite{natphys:trotzky}. 
}
\label{fig:HFSCFilling5n0vsT}
\end{figure}

 Within our Hartree-Fock approach interaction
effects arise because the boson density acts as an effective
spatially-varying single-particle potential.  At larger filling,
the boson density is higher and one may expect interaction effects to 
be stronger.  In the present section we illustrate this by increasing
the system filling to $f=5$ while keeping all other parameters
consistent with  Ref.~\cite{natphys:trotzky}.

Figure~\ref{fig:HFSCFilling5TcvsQ} shows the transition temperature as a function of
 normalized optical lattice depth for this case, showing a slight separation between 
the curves with increasing $q$.  Although $\tc$ is only slightly enhanced by interactions,
Fig.~\ref{fig:HFSCFilling5n0vsT}, which plots the condensate fraction below $\tc$ for the case of
$q=1.75$, shows a clear enhancement of the condensate fraction
 (solid curve) relative to the non-interacting case (dashed curve).  We interpret this,
physically, as being due to the fact that, as more bosons enter the condensate below $\tc$,
the spatially-inhomogeneous nature of the wave function leads, self-consistently, to a larger
effect of interactions on system properties.  At the lowest temperatures, however, all bosons
enter the condensate, so that both curves must eventually merge at $N_0/N = 1$ for $T\to 0$,
as seen in Fig.~\ref{fig:HFSCFilling5n0vsT}.

\begin{figure}[htb]
  \begin{center}
  \includegraphics[width=0.87\columnwidth]{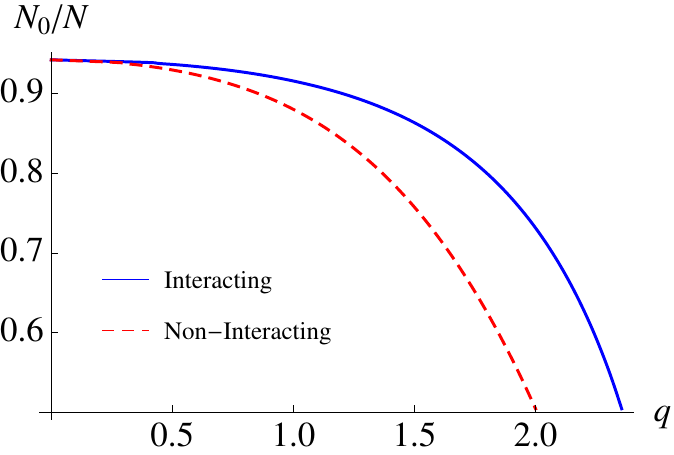}\\
  \end{center}
  \caption{(Color online) In this plot, we show the condensate fraction $N_0/N$
for both interacting and non-interacting gases, as a function of the
lattice depth $q=V_0/4E_{\rm r}$ at temperature $T=0.1E_{\rm r}$. Here the system is at unit
filling, with system parameters consistent with those of
Ref.~\cite{natphys:trotzky}, except with a large scattering length $a_s= 0.1a$. }
\label{fig:eight}
\end{figure}

\subsection{Interaction effects at large scattering length}

Next, we investigate the effect of increasing the $s$-wave scattering
length.  To characterize the interactions, we  note that, relative to
the lattice spacing, the Trotzky experiments are at $a_s/a\simeq
1.3\times 10^{-2}$, which we can regard as being at weak coupling.
However, larger scattering lengths are indeed achievable for cold
atoms in optical lattices, as shown, for example, by the experiments
of Mark et al.~\cite{prl:marknageri} on cesium BEC's.  To explore this,
we studied bosons in optical lattices within our self-consistent
approach, using the same parameters as the Trotzky et al.
experiments~\cite{natphys:trotzky} but with a larger scattering length
$a_s \simeq 0.1 a$.  As shown in Fig.~\ref{fig:eight}, this leads to a
rather large shift of the condensate fraction, relative to the
non-interacting case, that grows with increasing optical lattice
depth.

\begin{figure}[ht!]
     \begin{center}
        \subfigure{
            \includegraphics[width=85mm]{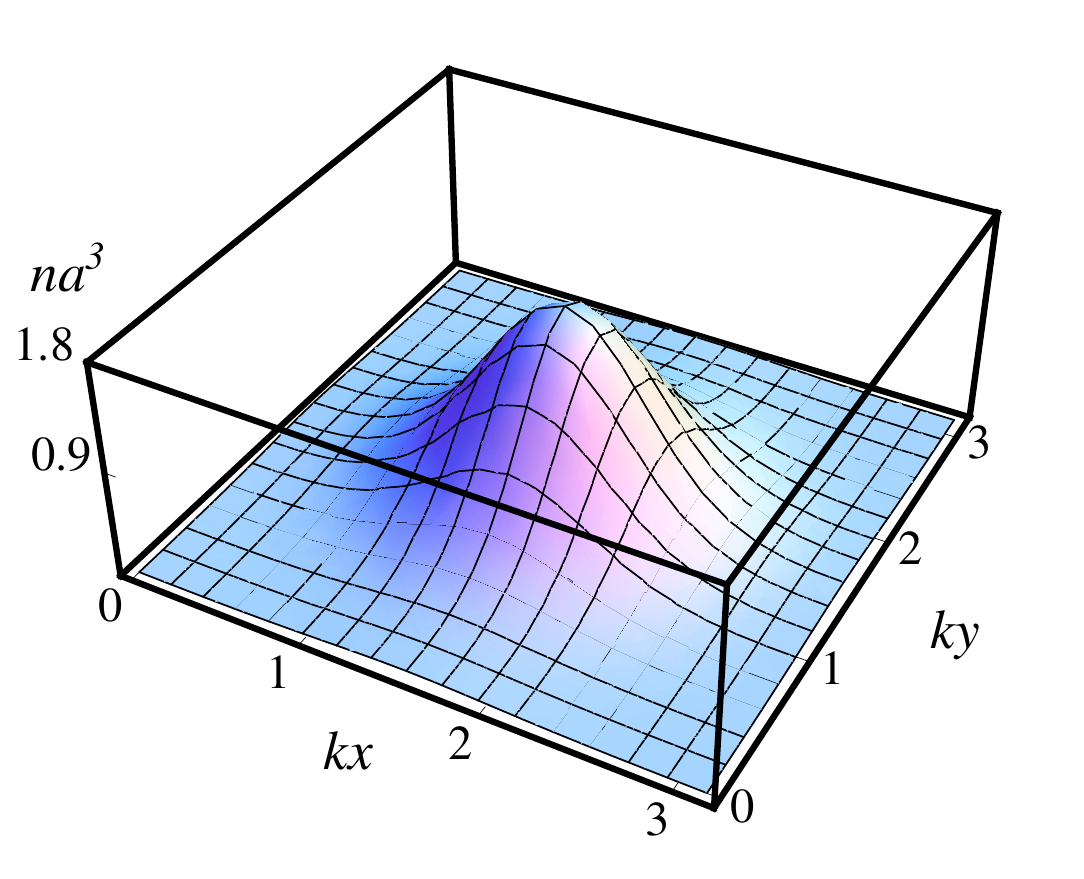}}
        \\ \vspace{-.25cm}
        \subfigure{
            \includegraphics[width=85mm]{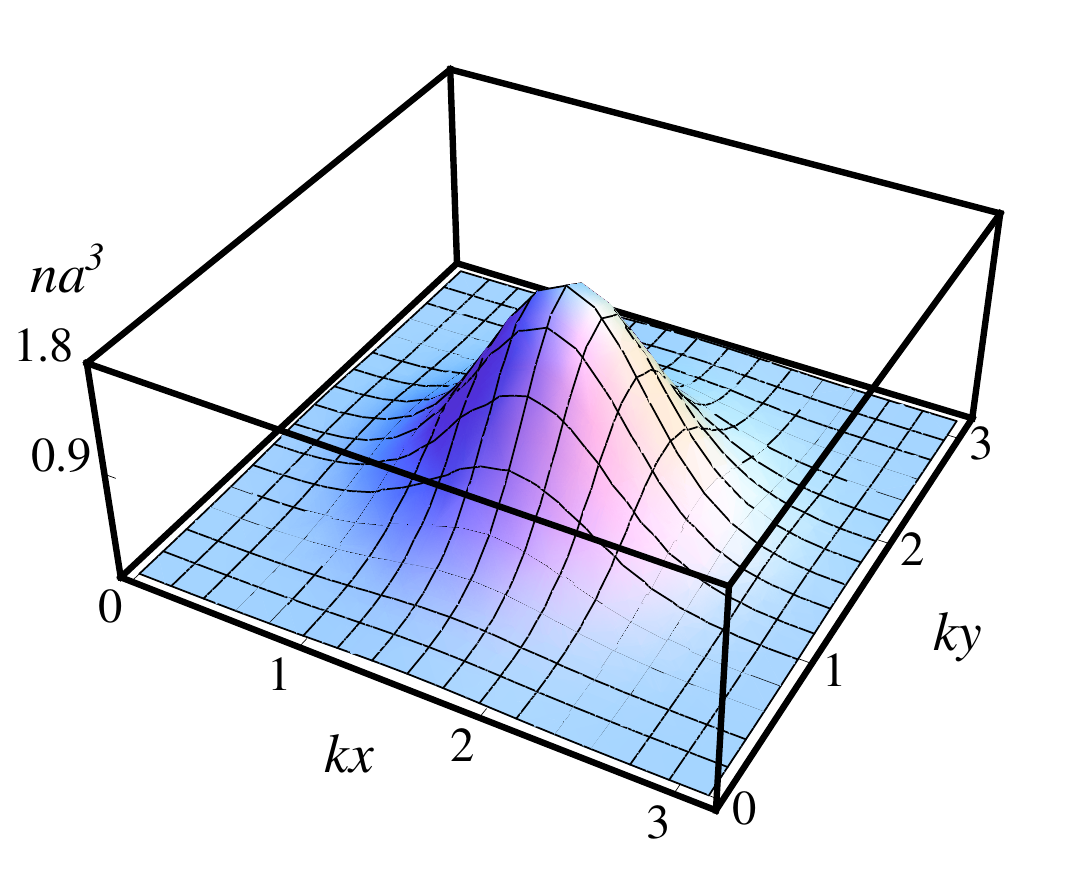}}
    \end{center}\vspace{-.5cm}
    \caption{ (Color online) The top panel shows the local density in a unit cell for $kz=\pi/2$ in the presence
of repulsive interactions.  Here, system parameter are the same as those of 
Ref.~\cite{natphys:trotzky} but with a larger scattering length $a_s \simeq 0.1 a$ and filling $f=2.83$.  For comparison,
the bottom panel shows the non-interacting case.
     }\vspace{-.25cm}
   \label{fig:LocalDenUnitCell}
\end{figure}

\subsection{Interaction effects on local boson density}

We have argued that the enhancement of the condensate fraction at a particular temperature and
optical lattice depth, relative to the non-interacting case, occurs because the local boson density is more spatially
uniform in the presence of repulsive interactions, enhancing superfluidity.  Computing the local density in a unit 
cell requires solving for our self consistent parameters $c$ and $\bar{q}$, along with the system chemical potential $\mu$
and then inputing these values into the local density Eq.~(\ref{equ:numberEqbelowTc}), requiring a sum over the 
Mathieu function indices $\gv{\nu} = (\nu_x,\nu_y,\nu_z)$.  After carrying out this numerically-intensive procedure, we 
find that the local density in a unit cell indeed becomes broadened within a unit cell with increasing repulsive 
interactions, as shown in Fig.~\ref{fig:LocalDenUnitCell} for $kz=\pi/2$.  Here the top panel is the interacting case, and the bottom panel
is the non-interacting case, with the system temperature given by  $\kb T/E_{\rm r} = 0.3846$, filling $f = 2.83$,
 optical lattice depth $q = 2$, and $a_s=0.1 a$.  For these parameters, the non-interacting plot is at $\tc$ while
the interacting plot is slightly below $\tc$.

\begin{figure}[htb]
  \begin{center}
  \includegraphics[width=0.97\columnwidth]{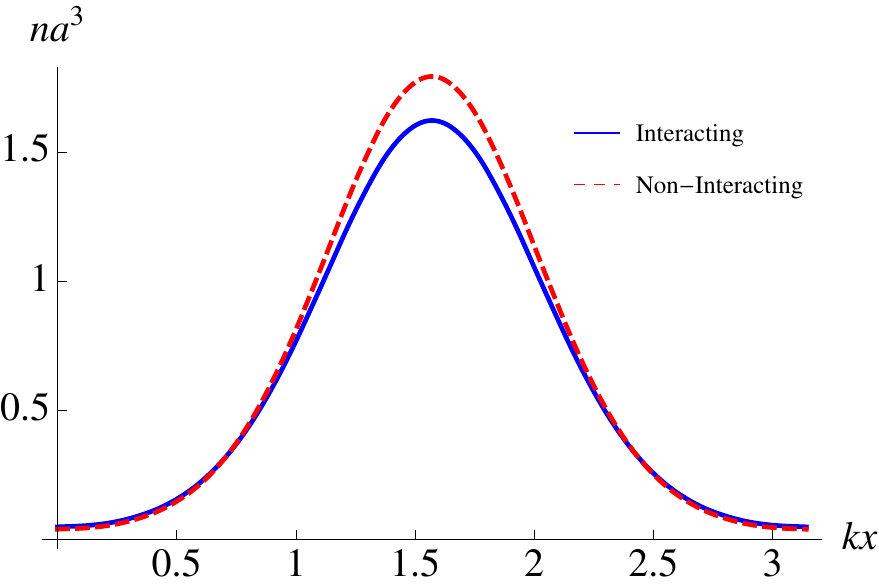}\\
  \end{center}
  \caption{(Color online) This plot shows the interacting boson density (solid curve) as a function of the
spatial variable $x$ in a unit cell at $ky=\pi/2$, $kz=\pi/2$ (crossing the unit cell center), showing a suppression of the central boson density 
relative to the non-interacting case (dashed curve).}\label{fig:RealDen1}
\end{figure}

\begin{figure}[htb]
  \begin{center}
  \includegraphics[width=0.97\columnwidth]{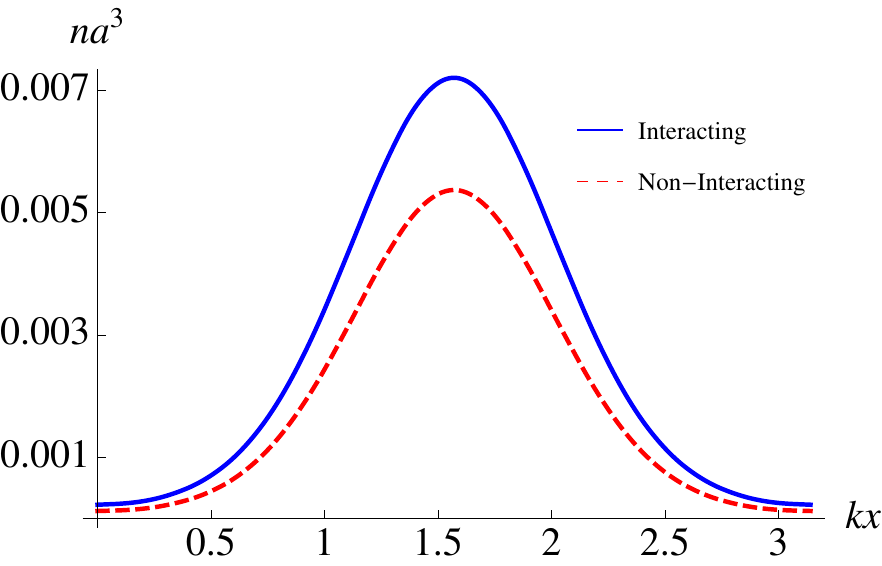}\\
  \end{center}
  \caption{
(Color online) This plot shows the interacting boson density (solid curve) as a function of the
spatial variable $x$ in a unit cell at $ky=\pi/8$, $kz=\pi/8$ (near the unit cell edge), showing an increase of the central boson density 
relative to the non-interacting case (dashed curve).
}\label{fig:RealDen2}
\end{figure}

For a more quantitative comparison, in Figs.~\ref{fig:RealDen1} and \ref{fig:RealDen2} we show 
the local density vs. position for  at the center $ky=\pi/2$,
$kz=\pi/2$ (Fig.~\ref{fig:RealDen1}) and near the edge $ky=\pi/8$, $kz=\pi/8$ (Fig.~\ref{fig:RealDen2}), showing that the boson density
is more homogeneous in the interacting case, with the interacting density
smaller near the unit cell center and larger at the edge of the unit cell, relative to the non-interacting case.

\subsection{Boson momentum distribution}

In ultracold atom experiments, a BEC is indicated by peaks (i.e.
maxima) in the images of the cloud after free expansion that reflect the boson momentum distribution
in the initially trapped cloud.  In this section, we calculate the momentum distribution
to see how interaction changes the superfluid state.  As reviewed in Appendix~\ref{sec:freeexp},
the real space boson density after free expansion probes the momentum distribution: 
\begin{equation}
n(\mathbf{k})=N_0 | \Phi_0(\mathbf{k})|^2 + \sum_{i\neq 0}  | \Phi_i(\mathbf{k})|^2
\nb(\epsilon_i-\mu),
\label{equ:totaldensity}
\end{equation}
 where $\Phi(\mathbf{k})$ is the Fourier transformed
Mathieu wave function. By inserting the wave functions and energy
levels obtained from our self-consistent scheme into
Eq.~\eqref{equ:totaldensity}, we are able to obtain the boson momentum distribution. Note that we are not considering
interaction during the expansion.

\begin{figure}[ht!]
     \begin{center}
        \subfigure{
            \includegraphics[width=85mm]{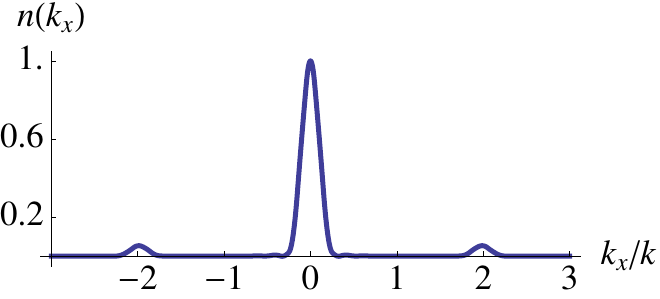}}
        \\ \vspace{-.25cm}
        \subfigure{
            \includegraphics[width=85mm]{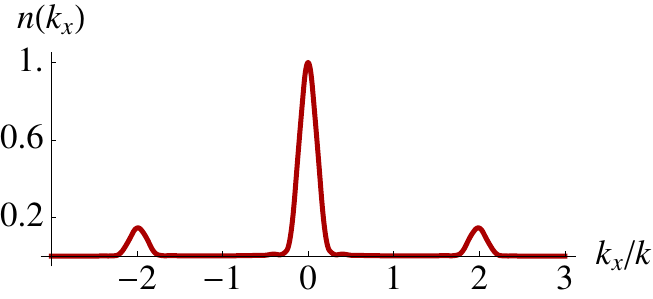}}
    \end{center}\vspace{-.5cm}
    \caption{(Color online) The boson momentum distribution as a function of $k_x$ for $k_y = 0$ and $k_z=0$, with system parameters given
in the main text.  Both plots are normalized so that $n(0) = 1$.
  The top panel shows the interacting case, and the bottom panel shows the non-interacting case, with the 
height of the side peaks being smaller in the interacting case reflecting a more spatially-uniform Bose gas.  
     }\vspace{-.25cm}
   \label{fig:ExpandDens1D}
\end{figure}

Most cold atom experiments with optical lattices involve a 
background smoothly-varying parabolic trap.  In our analysis, we did 
not account for this, but instead studied a \lq\lq box\rq\rq-shaped trap
possessing a periodic optical lattice potential along with hard-wall
boundary conditions to take into account the finite-size initial cloud.  
We considered a cubic system with length $L = \pi N_{\rm s}/k$ with $N_{\rm s} = 10$ lattice
sites along each direction (containing $\nsites = 10^3$ total sites).   
We note here that such box-shaped traps have recently been achieved experimentally~\cite{prl:gaunt}.

In Fig.~\ref{fig:ExpandDens1D}, we show our results for the boson momentum distribution Eq.~(\ref{equ:totaldensity})
at $k_y=k_x = 0$, with the same system parameters as in the preceding section, with the top panel
being the interacting case, and the bottom panel being the non-interacting case.  Each plot
is normalized so that $n(\bk) = 1$ at $\bk=0$.
 The side-peaks are expected
for a BEC in a periodic optical lattice (as observed by Greiner et al.~\cite{nat:greiner}), 
and should occur for $\bk$ equal to any reciprocal lattice
vector.  The side-peaks in Fig.~\ref{fig:ExpandDens1D} occur at  $k_x = \pm \frac{2\pi}{a} = \pm \frac{2}{k}$.
The height of the side peaks, relative to the central peak, reflects the degree of spatial
inhomogeneity of the BEC, as can be seen by noting the limiting case of a spatially-uniform BEC, which will have
only a central peak at $\bk = 0$.  Thus, since the side-peaks are smaller in the interacting case,
we argue that the boson momentum distribution also reflects the spatial homogenization of the cloud
in the presence of repulsive interactions.

\section{Conclusion}
\label{sec:conclude}

In this paper, we studied the effect of short range repulsive interactions on 
the properties of bosons in periodic optical lattices.  For a uniform boson
gas, the effect of such interactions on the superfluid transition temperature
$\tc$ has been argued for decades, with both positive and negative $\tc$ shifts
having been reported~\cite{rmp:andersen}.  The consensus is that $\tc$ increases
linearly with scattering length $\as$~\cite{Baym99}. 

In contrast, bosons in a deep optical lattice, characterized by the Bose Hubbard model (BHM),
exhibit a suppression of $\tc$ with increasing repulsive interactions.  Given that
a system consisting of bosons in a periodic optical lattice  with lattice depth $V_0$ 
continuously interpolates between the limiting cases of a uniform gas (for $V_0 \to 0$)
and the BHM (for $V_0\gg E_r$), then one may expect an increase of $\tc$ with $\as$ 
away from the BHM regime.

Instead of the commonly used tight-binding approach that leads to the BHM, 
our theoretical study of this system  started from
the exact single-particle states of bosons in an optical lattice,
satisfying the  Mathieu equation, an approach that can be particularly
useful at large boson filling or when many single-particle bands are occupied.  
Interaction effects were accounted for using a self-consistent Hartree-Fock 
approximation, in which the spatially-inhomogeneous boson density leads to an
effective reduction of the optical lattice depth.

We applied this
scheme to quantify the effects of inter-atomic interactions on the
properties of bosons in an optical lattice, as exhibited in the
comparison between observables of non-interacting and interacting
systems. We found that interactions increase the superfluid transition temperature and the condensate
fraction, and also homogenizes the local boson density (as would be seen in the local density and
also the momentum distribution as probed in time-of-flight experiments).

An obvious weakness of our approach is that we are unable to capture the Mott 
insulating regime for bosons in optical lattices occurring for integer filling 
at large optical lattice depth.  A natural extension of our work will be to
understand the emergence of the Mott insulating phase within the Mathieu
equation approach (i.e., without making the BHM approximation which provides a natural
picture of the Mott insulating state).  Such an extension would lead to a more complete understanding
of the properties of interacting BEC's in optical lattices.  

This work was supported by National Science Foundation Grant No. DMR-1151717.
This work was supported in part by the National Science Foundation under 
Grant No. PHYS-1066293 and the hospitality of the Aspen Center for Physics.
DES acknowledges support from the
German Academic Exchange Service (DAAD) and the hospitality of the 
Institute for Theoretical Condensed Matter physics at the Karlsruhe Institute of 
Technology.

\appendix

\section{Hartree-Fock Approximation}
\label{appnd:HF}
We use Hartree-Fock Approximation to describe the inter-atomic interaction. Hartree-Fock approximation assumes a dependence of the system's self energy on the atomic density.
 
Consider a translationally invariant system or a system possessing discrete translational invariance (such as in a periodic potential). The Dyson's Equation is
\begin{widetext}
\begin{equation}
G(\mathbf{r},\mathbf{r}^\prime;\tau, \tau^\prime) 
 =G_0(\mathbf{r},\mathbf{r}^\prime;\tau, \tau^\prime)+\int
d^3 r_1 d^3 r_2 d\tau_1 d\tau_2 ~G_0(\mathbf{r},{\mathbf{r}_1}^\prime;\tau, \tau_1)\Sigma(\mathbf{r}_1,\mathbf{r}_2;\tau_1, \tau_2)G(\mathbf{r}_2,\mathbf{r}^\prime;\tau_2, \tau^\prime),
\label{equ:dysonEq}
\end{equation}
\end{widetext}
where $G$ and $G_0$ are the Green's functions of spatial coordinates $\br$, $\br^\prime$ and imaginary time $\tau$, $\tau^\prime$ respectively for the entire system and for the bare system, and $\Sigma$ is the self energy characterizing the contribution from interaction. On the other hand, we have
\be
G(\br,\br^\prime,\tau,\tau^\prime)=-\avg{T_\tau \Psi(\br,\tau)\Psi^\dagger(\br^\prime,\tau^\prime) e^{-\int_0^\beta d\tau H_1(\tau)}},
\ee
where $H_1$ is given by Eq.~(\ref{eq:hone}). To derive the Hartree-Fock term, we expand to the first order (denoting $x=(\br,\tau)$, and similarly for $x^\prime$ and $x_1$),
\be
\begin{split}
&G(x,x^\prime)=\\
&-\avg{T_\tau \Psi(x)\Psi^\dagger(x^\prime)\Big[1-\f{g}{2}\int dx_1~\Psi^\dagger(x_1)\Psi^\dagger(x_1) \Psi(x_1) \Psi(x_1)\Big]},
\end{split}
\ee
where the coupling constant $g=4\pi \hbar^2 a_s/m$. In the above Green's function, there are two ways for $\Psi(x)$ to contract with the two $\Psi^\dagger(x_1)$ factors, and there are two ways for $\Psi^\dagger(x^\prime)$ to contract with the two $\Psi(x_1)$ factors. Therefore, with $G_0(x,x^\prime)=-\avg{T_\tau \Psi(x)\Psi^\dagger(x^\prime)}$, we have
\be
G(x,x^\prime)=G_0(x,x^\prime)-2g\int dx_1~G_0(x,x_1) G_0(x_1, x_1) G_0(x_1,x^\prime).
\label{equ:perturbexpand}
\ee
Comparing Eq.~(\ref{equ:perturbexpand}) with Eq.~(\ref{equ:dysonEq}), we obtain
\be
\Sigma(x_1, x_2)=-2gG_0(x_1, x_1) \delta(x_1-x_2).
\ee
Since $G_0(x_1, x^{+}_1) =-n(x_1)$, the boson density, we come to 
\be
\Sigma(x, x^\prime)=2g n(x) \delta(x-x^\prime),
\ee
namely, the Hartree-Fock Approximation.

\section{Free Expansion}
\label{sec:freeexp} The superfluid state of bosons is usually shown in
experiments by absorption imaging of the freely expanded cloud~\cite{read}. In
this process, the trapping potential is abruptly turned off, the
atomic gas undergoes a period of time-of-flight free expansion. The
absorption image provides information about the density profile, which
is related to the initial momentum distribution. Atoms that are
initially of the same state will gather together in the real space,
therefore density peaks in the image indicate Bose-Einstein
condensate.

Assume the trapping potential is turned off at $t=0$. The time dependent density
\be
n(\mathbf{r},t)=\mathrm{Tr}\rho_{\mathcal{H}}(t)\hat{n}(\mathbf{r})=\mathrm{Tr}\rho_{\mathcal{H}}(t)\hat{\Phi}^\dagger(\mathbf{r})\hat{\Phi}(\mathbf{r}),
\ee
where $\hat{\Phi}(\mathbf{r})$ and $\hat{\Phi}^\dagger(\mathbf{r})$ are field operators, $\rho_{\mathcal{H}}$ is the density matrix for the time dependent Hamiltonian which is written as
\be
\mathcal{H}(t)=H_0+\Theta(-t)H_{\rm trap},
\ee
where
\be
\Theta(t)=\left\{\begin{array}{l}
0,~~(t\leq 0);\\
1,~~(t>0).
\end{array}\right.
\ee
The Hamiltonian is entirely time invariant before $t=0$, we can denote $\mathcal{H}(t<0)\equiv \mathcal{H}_\infty$. The density after the trap potential is turned off is obtained by
\begin{equation}
n(\mathbf{r},t>0)=\mathrm{Tr}\rho_{\mathcal{H}_\infty}\hat{n}(\mathbf{r},t),
\label{equ:expden}
\end{equation}
where $\hat{n}(\mathbf{r},t)\equiv  e^{itH_0} {n}(\mathbf{r})e^{-itH_0}$. $\rho_{\mathcal{H}_\infty}\equiv\rho_{\mathcal{H}}(t<0)$ is the initial density matrix at $t=0$.
The field operators $\hat{\Phi}(\mathbf{r})$ can be expanded either in the eigenstates of $H_0$ or in the eigenstates of $\mathcal{H}_\infty$, and the former of which are just plane waves. Therefore we can obtain the density matrix $\rho_{\mathcal{H}_\infty}$ by equating the two expansions, and plug into Eq.~(\ref{equ:expden}) to solve for the density $n(\mathbf{r},t>0)$. The result is
\be
n(\mathbf{r},t)=\sum_{i}\sum_{\mathbf{k}}\mid e^{itE_{\mathbf{k}}-i\mathbf{k}\cdot\mathbf{r}}\tilde{\phi}_{i}(\mathbf{k})\mid^2 \nb(\varepsilon_{i}-\mu),
\label{equ:freeexpdensitymaintext}
\ee
where $\tilde{\phi}_{i}(\mathbf{k})$ is the Fourier transform of the eigenfunction of $\mathcal{H}_\infty$. If we translationally move in the momentum space from $\mathbf{k}$ to $\mathbf{k}+\f{m\mathbf{r}}{t}$, where $\mathbf{k}\ll\f{m\mathbf{r}}{t}$, Eq.~(\ref{equ:freeexpdensitymaintext}) becomes
\begin{equation}
n(\mathbf{r},t)\propto \sum_{i}\abs{ \tilde{\phi}_{i}(\mathbf{r}m/t)}^2 \nb(\varepsilon_{i}-\mu).
\label{equ:freeexpdenmaintext}
\end{equation}
We note that, up to an overall prefactor, the right side of Eq.~(\ref{equ:freeexpdenmaintext}) is the momentum distribution $n(\mathbf{k})$ of the initial trapped gas, measured at momentum $\mathbf{k}=m\mathbf{r}/t$. Thus, the density profile of the expanded cloud allows experimentalists to probe the momentum distribution of the trapped cloud.


\begin{thebibliography}{10}
\bibitem{rmp:bloch} I. Bloch, J. Dalibard, and W. Zwerger, Rev. Mod. Phys. {\bf 80}, 885
(2008).
\bibitem{review:Lewenstein} M. Lewenstein, A. Sanpera, V. Ahufinger, B. Damski, A. Sen(De) and U. Sen, Adv. Phys. \textbf{56}, 243 (2007).


\bibitem{prl:jaksch1998}D. Jaksch, C. Bruder, J. I. Cirac, C. W. Gardiner, and P. Zoller, Phys. Rev. Lett. {\bf 81}, 3108 (1998).
\bibitem{prb:fisher1989} M. P. A. Fisher, P. B. Weichman, G. Grinstein, and D. S. Fisher,
Phys. Rev. B {\bf 40}, 546 (1989).

\bibitem{nat:greiner} M. Greiner, O. Mandel, T. Esslinger, T. W. H\"ansch, and I. Bloch, {Nature} \textbf{415}, 39 (2002).

\bibitem{jlowtemp:kohl}M. K\"ohl, H. Moritz, T. St\"oferle, C. Schori, and T. Esslinger, J. Low Temp. Phys. {\bf 138}, 635 (2005).

\bibitem{prl:spielman2007}I. B. Spielman, W. D. Phillips, and J. V. Porto, Phys. Rev. Lett. {\bf 98}, 080404 (2007).

\bibitem{prl:MunKetterle}J. Mun, P. Medley, G. K. Campbell, L. G. Marcassa, D. E. Pritchard, and W. Ketterle, Phys. Rev. Lett. {\bf 99}, 150604 (2007).
\bibitem{nat:chin}N. Gemelke, X. Zhang, C.-L. Hung, and C. Chin, Nature {\bf 460}, 995 (2009). 
\bibitem{natphys:trotzky} S. Trotzky, L. Pollet, F. Gerbier, U. Schnorrberger, I. Bloch, N. V. Prokof'ev, 
B. Svistunov, and M. Troyer, {Nature Physics} \textbf{6}, 998 (2010).



\bibitem{njp:becker}C. Becker, P. Soltan-Panahi, J. Kronj\"ager, S. D\"orscher, K Bongs, and K Sengstock, 
New J. Phys. {\bf 12}, 065025 (2010).
%
\bibitem{sci:endres2011}M. Endres, M. Cheneau, T. Fukuhara, C. Weitenberg, P. Schau\ss, C. Gross, L. Mazza, 
M. C. Ba\~nuls, L. Pollet, I. Bloch, and S. Kuhr, Science {\bf 334}, 200 (2011).
%
\bibitem{prl:marknageri} M. J. Mark, E. Haller, K. Lauber, J. G. Danzl, A. J. Daley, and H.-C. N\"agerl,
 Phys. Rev. Lett. {\bf 107}, 175301 (2011).
%
\bibitem{sci:chin} X. Zhang, C.-L. Hung, S.-K. Tung, and C. Chin, Science {\bf 335}, 1070 (2012). 








\bibitem{Capogrosso}
B. Capogrosso-Sansone, N.V. Prokof'ev, and B.V. Svistunov, Phys. Rev. B {\bf 75}, 134302 (2007),
\bibitem{Abramowitz} {\em Handbook of Mathematical
Functions}, edited by M. Abramowitz and I. A. Stegun (Dover, New York,
1972).
\bibitem{Zwerger} W. Zwerger, 
J. Opt. B: Quantum and Semiclassical Optics {\bf 5}, S9 (2003). 
\bibitem{pra:demarco}D. McKay, M. White, and B. DeMarco, Phys. Rev. A {\bf 79}, 063605 (2009).
\bibitem{Baym99} G. Baym, J.-P. Blaizot, M. Holzmann, F. Lalo\"e, and D. Vautherin, Phys. Rev. Lett. {\bf 83}, 1703 (1999).

\bibitem{rmp:andersen} J. O. Andersen, {Rev. Mod. Phys.} \textbf{76}, 599 (2004).

\bibitem{web:nistmath}NIST Digital Library of Mathematical Functions, http://dlmf.nist.gov/ .





\bibitem{prl:gaunt} A.L. Gaunt, T. F. Schmidutz, I. Gotlibovych, R. P. Smith, and Z. Hadzibabic,  Phys. Rev. Lett. {\bf 110}, 200406 (2013).


\bibitem{read}
N. Read and N.R. Cooper, Phys. Rev. A {\bf 68\/}, 035601 (2003).

\end{thebibliography}
\end{document}